\documentclass[runningheads]{llncs}
\usepackage{graphicx}
\usepackage{hyperref,xcolor}
\hypersetup{
    colorlinks = true,
    linkcolor = {blue},
    citecolor = {blue},
}

\usepackage{mathtools}
\usepackage{listings}
\usepackage{proof}
\usepackage{amsfonts}
\usepackage{amssymb}
\usepackage{stmaryrd}

\usepackage[misc]{ifsym}

\begin{document}
\title{Formalizing and Verifying Decentralized Systems with Extended Concurrent Separation Logic
}
\titlerunning{Formalizing and Verifying Decentralized Systems with Extended CSL}
%
\author{Yepeng Ding \and
Hiroyuki Sato}
\authorrunning{Y. Ding and H. Sato}
%
\institute{The University of Tokyo, Tokyo, Japan\\
\email{\{youhoutei,schuko\}@satolab.itc.u-tokyo.ac.jp}}
\maketitle              
\begin{abstract}
Decentralized techniques are becoming crucial and ubiquitous with the rapid advancement of distributed ledger technologies such as the blockchain. Numerous decentralized systems have been developed to address security and privacy issues with great dependability and reliability via these techniques. Meanwhile, formalization and verification of the decentralized systems is the key to ensuring correctness of the design and security properties of the implementation. In this paper, we propose a novel method of formalizing and verifying decentralized systems with a kind of extended concurrent separation logic. Our logic extends the standard concurrent separation logic with new features including communication encapsulation, environment perception, and node-level reasoning, which enhances modularity and expressiveness. Besides, we develop our logic with unitarity and compatibility to facilitate implementation. Furthermore, we demonstrate the effectiveness and versatility of our method by applying our logic to formalize and verify critical techniques in decentralized systems including the consensus mechanism and the smart contract.

\end{abstract}
\section{Introduction}
Nowadays, decentralized technology has evolved into a new stage with the advancement of many kinds of decentralized techniques such as the consensus mechanism, smart contract. Based on these decentralized techniques, numerous solutions have been proposed to circumvent security and privacy issues widely existing in centralized systems such as unauthorized disclosure, deception, disruption, and usurpation, which is attracting a huge amount of attention from both academy and industry. These solutions have been developed as decentralized systems and applied to a wide range of fields such as economics \cite{bohme_bitcoin_2015}, the Internet of things \cite{ding_bloccess_2020}, smart health \cite{ding_derepo_2020}, and infrastructures \cite{ding_dagbase_2020}. However, it is challenging to ensure the correctness and properties of decentralized techniques and systems due to the high complexity of the design and intricate factors.

Formal methods play an important role in verifying complex systems by specifying systems with rigorous mathematical syntax and semantics to eliminate imprecision and ambiguity of the design and implementation. Significant contributions have been made to verify distributed systems with two mainstream techniques of formal methods including model checking and automated theorem proving. Model-checking techniques \cite{konnov_completeness_2017,fatkina_methods_2019,souri_symbolic_2019} prove to be effective to check safety and liveness properties of distributed system design and algorithms. However, it is noteworthy that these model checking techniques suffer from the state-space explosion problem, which leads to the ineffectiveness of complicated realistic systems, though many model-checking tools such as SPIN \cite{holzmann_model_1997}, NuSMV \cite{cimatti_nusmv_2002}, Cubicle \cite{conchon_cubicle_2012} provide space-efficient and on-the-fly algorithm to optimize the methods. Meanwhile, theorem proving techniques \cite{hawblitzel_ironfleet_2015,rahli_formal_2015,sardar_theorem_2017} provide a deductive method to prove the correctness and properties of distributed systems in a mathematical style without the requirement of the exploration of the state space. To date, the proof assistant is the main tool for the development of formal proofs with the collaboration of humans and machines such as Coq \cite{noauthor_welcome_nodate} and Isabelle \cite{noauthor_isabelle_nodate}. These formal methods make it possible to locate design flaws and implementation pitfalls and ensure correctness and properties. Particularly, sound logic is imperative in formal reasoning about program correctness. Concurrent separation logic (CSL) \cite{ohearn_resources_2007} has been widely used to reason about concurrent systems such as cryptographic implementation \cite{appel_verification_2015}, the concurrent operating system kernels \cite{xu_practical_2016}. As the extension of separation logic \cite{ohearn_logic_1999,ishtiaq_bi_2001}, CSL enhances the modularity to reason about resources locally, which makes it capable of formalizing the disjoint concurrency and inter-process interactions.

Admittedly, it is natural that decentralized technology highly requires the guarantee of correctness and system properties since there is no central entity to supervise and monitor network and system behaviors, especially in untrustworthy environments. Nevertheless, formal reasoning about decentralized systems with the standard CSL is restrictive.

\begin{itemize}
\item CSL lacks encapsulated components to reason about the communication that is the basic and non-negligible component in the specification of decentralized systems to formalize interactions.
\item CSL has restrictive expressiveness of reasoning about the environment factor and temporal conditions that can facilitate formalizing complex protocols of decentralized systems.
\item CSL focuses on reasoning about low-level programs such as memory management and resource relationship. It is restrictive for CSL to reason about high-level systems such as nodes in decentralized systems.
\end{itemize}

Hence, it is still a significant challenge to reason about decentralized systems with an effective logic with rich expressiveness and high modularity.

In this paper, we propose a method of formalizing and verifying decentralized systems with an extended concurrent separation logic with three novel features: communication encapsulation, environment perception, and node-level reasoning. These features are developed to bridge the gulf of reasoning about decentralized systems with CSL. We encapsulate the formalization of communications as the basic specification and proof component. Besides, our logic enriches the expressiveness of CSL with the capability of environment perception. The environment perception also extends CSL into two-dimension reasoning including both spatial and temporal reasoning. Moreover, we introduce the node-level reasoning to enable our logic to formalize high-level decentralized systems, which enhances the modularity to specify and verify systems at different levels.

We summarize our main contributions as follows:
\begin{enumerate}
\item We propose a novel method of formalizing and verifying decentralized systems with a kind of extended concurrent separation logic. Our logic addresses the issues of CSL while reasoning about decentralized systems with novel features including communication encapsulation, environment perception, and node-level reasoning.
\item We formalize a consensus mechanism with our method to prove the effectiveness of reasoning about complex protocols in decentralized systems with our logic. Our logic simplifies high-level reasoning with great modularity and rich expressiveness. It also presents unitarity while reasoning about systems at different abstraction levels.
\item We also demonstrate our work by applying our logic to the specification and verification of smart contracts. We locate the design flaw of a typical smart contract with vulnerability to prove the effectiveness of formalizing and verifying decentralized applications.
\end{enumerate}

The remainder of this paper is organized as follows. We give a short introduction to the related work in Section~\ref{sec:related}. We then illustrate our logic that is the core of our method in Section~\ref{sec:logic}. Subsequently, we describe the method of the application of our logic in a consensus algorithm and a smart contract in Section~\ref{sec:app}. We discuss our work in Section~\ref{sec:discuss}, followed by the conclusion in Section~\ref{sec:conclusion}.

\section{Related work}
\label{sec:related}
Prior work has made significant contributions to the formalization and verification of distributed systems such as \cite{sardar_theorem_2017,aminof_parameterized_2018,fatkina_methods_2019,souri_symbolic_2019}. Model checking and theorem proving are two main methods in these works with the support of typical tools such as HOL4 \cite{gordon_introduction_1993}, TLC model checker \cite{yu_model_1999}, NuSMV \cite{cimatti_nusmv_2002}. In addition, reasoning about distributed systems with Hoare-style logic has also proved effective. IronFleet \cite{hawblitzel_ironfleet_2015} was proposed to build practical and provably correct distributed systems based on the blend of TLA-style state-machine refinement and Hoare-logic verification. DISEL \cite{sergey_programming_2017} provided a framework for implementation and compositional verification of distributed systems and clients based on the distributed separation logic. Particularly, our logic improves the work \cite{ding2020extending} to make it practical to formalize and verify the decentralized systems.

In the meanwhile, the decentralized technology has been rapidly developed to address the common issues in centralized distributed systems since Bitcoin \cite{raval_decentralized_2016}. Due to the lack of supervision from central entities, ensuring the correctness and properties of decentralized systems during the development is imperative. Formal verification provides a mathematical approach to analyze the decentralized system in a rigorous manner. However, the specific research of the formalization and verification of decentralized systems just gets started with the boost of decentralized technology.

As one of the critical techniques in decentralized systems, the consensus algorithm has been applied with formal methods to ensure the correctness in the distributed environment. The agreement safety property of the PBFT \cite{castro_practical_1999} was proved in \cite{rahli_velisarios_2018}. Besides, the Raft \cite{ongaro_search_2014} state replication library was formally verified by Verdi \cite{wilcox_verdi_2015}, a framework for formal verification of distributed systems implemented in Coq.

In recent years, formal verification of smart contracts has been an attractive topic since TheDAO attack \cite{atzei_survey_2017} that brought great damage to the cryptocurrency market and successfully transferred about \$50M worth of Ether into the control of the attacker by exploiting the reentrancy vulnerability. In the prompt work \cite{bhargavan_formal_2016}, they proposed a method to translate smart contracts implemented in Solidity to F* \cite{swamy_dependent_2016} and decompile Ethereum virtual machine (EVM) bytecode into F* for formal verification. Later, model-checking-based approaches were proposed for formal verification of smart contracts. In the work \cite{bai_formal_2018}, SPIN has been used to verify the correctness and properties of a smart contract template to reduce the potential errors.

The EVM that supports the execution of smart contracts has also been formally specified and verified. In \cite{hildenbrandt_kevm_2018}, the formal specification of the EVM bytecode named KEVM was developed with $\mathbb{K}$ framework \cite{rosu_overview_2010}, which provides the foundations for the verification. A deductive verifier was constructed in \cite{park_formal_2018} to precisely reason about possible behaviors of the EVM bytecode with KEVM.

\section{Our Logic}
\label{sec:logic}
The standard CSL has great significance in reasoning about concurrent programs with preeminent expressiveness. It is noteworthy that the standard CSL and typical variants can well support the formalization of parallel systems at the thread level or process level about memory management and resources. Our logic is compatible with CSL and inherits the capability of formalizing low-level systems. Based on that, we extend the standard CSL into formalizing high-level decentralized systems with better modularity and richer expressiveness.

\subsection{Communication Encapsulation}
In a decentralized system, communications among nodes are indispensable. It is hard for CSL to formalize the complex interactions among programs in an elegant manner. Hence, we simplify the formalization by encapsulating communications as the basic component.

Our logic defines a minimal program unit over $(\textit{Var}, \textit{Ch})$ in \eqref{f:program_unit}.

\begin{align}
\label{f:program_unit}
\mathfrak{P} \triangleq (L, A, \mathcal{E}, \hookrightarrow, L_0, g_0)
\end{align}

Here, $\textit{Var}$ is a set of typed variables and $\textit{Ch}$ is a set of channels. $L$ is a set of locations and $A$ is a set of actions. $\mathcal{E}$ denotes the effect function $A \times \llbracket \textit{Var\,} \rrbracket \mapsto \llbracket \textit{Var\,} \rrbracket$. The notation ${\hookrightarrow} \subseteq L \times \| \textit{Var\,} \| \times A \times L$ represents the conditional transition relation. $L_0 \subseteq L$ and $g_0 \in \| \textit{Var\,} \|$ denotes a set of initial locations and the initial condition respectively. $\llbracket \textit{Var\,} \rrbracket$ denotes the set of variable evaluations. $\| \textit{Var\,} \|$ denotes the set of Boolean conditions over $\textit{Var\,}$.

For convenience, we use the notation $l \xhookrightarrow{g:\alpha} l'$ as shorthand for $(l,g,\alpha,l') \in \hookrightarrow$ where $l \in L$ and $\alpha \in A$, meaning that the program $\mathfrak{P}$ goes from location $l$ to $l'$ when the current variable evaluation $\eta \models g$. We connect the program unit with CSL by specifying $l \xhookrightarrow{g:\alpha} l'$ as $\{ g \}~\alpha~\{ g' \}$, where $\mathcal{E}(\alpha,\eta) \models g'$.

Let $c!s$ denote sending signal $s$ via channel $c$ and $c?v$ denote receiving a signal from channel $c$ and assign the signal to variable $v$. A communication $\pi \in \Pi = \{ c!s, c?v \}$ is an action where $c \in \textit{Ch}, s \in \textit{Dom}(c), v \in \textit{Var} \text{ with } \textit{Dom}(v) \supseteq \textit{Dom}(c)$.

In a practical communication scenario, the finite asynchronous channel is commonly used. The essence of a finite asynchronous channel is a buffer with a capacity $\textit{Cap}(c) \in \mathbb{N}^{+}$ and a domain $\textit{Dom}(c)$. In this manner, a communication $c!s$ produces signal $s$ into the buffer whereas a communication $c?v$ consumes a signal from the buffer while assigning it to variable $v$. 

For example, two parallel programs $c!s$ and $c?v$ can be specified at a high specification level in \eqref{f:eg_comm_spec}.

\begin{align}
\label{f:eg_comm_spec}
\{ s \mapsto - \}~c!s \parallel c?v~\{ \llbracket v \rrbracket = \llbracket s \rrbracket \}
\end{align}

\begin{figure}
\centering
\includegraphics[width=3.8cm]{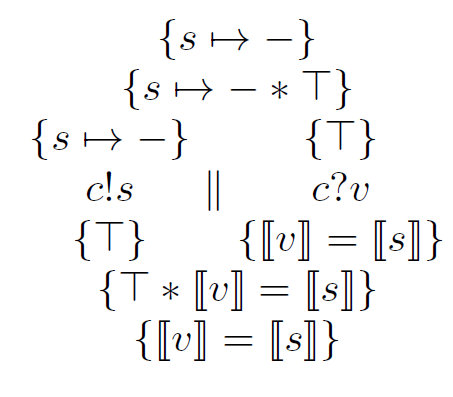}
\caption{Proof outline of $\{ s \mapsto - \}~c!s \parallel c?v~\{ \llbracket v \rrbracket = \llbracket s \rrbracket \}$.}
\label{fig:com_proof}
\end{figure}

We can give a proof outline of this parallel system containing the communication between two parallel programs in Fig.~\ref{fig:com_proof}.

Since we encapsulate the formalization of communications as the basic component, we have $\Pi \subseteq A$. The specification of communications can be automatically verified in the manner of Fig.~\ref{fig:com_proof}.

\subsection{Environment Perception}
\label{sec:env}
Our logic has the capability of perceiving the environment factors including foreign factor and native factor, which means that we can formulate specifications with both internal (native) and external (foreign) temporal conditions.

In our logic, we follow the style of Hoare triples as \eqref{f:style}.
\begin{align}
\label{f:style}
    \{ \Gamma, \gamma \land P \}~\alpha~\{ \Gamma, \gamma' \land P' \}
\end{align}

Here, $\Gamma$ is used to specify the foreign pre-conditions and post-conditions while $\gamma$ and $\gamma'$ are used to specify the native pre-conditions and post-conditions. $P$ and $P'$ are assertions with the same semantics of the standard CSL. $\alpha \in A$ is the action to change the state of programs.

Before illustrating the structure of the environment factor, we firstly introduce a partially ordered relation $\triangleleft$ defined in \eqref{f:occur}.

\begin{align}
\label{f:occur}
a \triangleleft a' \iff a=\textit{Pred}(a')
\end{align}

Here, $a,a' \in \acute{A}$ and $\textit{Pred}(a)$ denotes the predecessor action set of $a$. $\acute{A}$ is the set of occurred actions derived from a partial function $A \rightharpoonup \acute{A}$.

We use the notation $a \triangleleft a'$ as shorthand for $(a,a') \in \triangleleft$. Intuitively, $a \triangleleft a'$ means that action $a$ happens before action $a'$. Furthermore, the relation $\triangleleft$ has transitivity that is $a \triangleleft a' \triangleleft a'' \implies a \triangleleft a''$. With this ordered relation, we define a finite action path $\varrho$ as a finite action sequence $a_0 a_1 ... a_n$ such that $\forall i \in [0,n): (a_i,a_{i+1}) \in \triangleleft$, where $n \geq 1$ if the length of the sequence is greater than 1.

Here, we use the finite action path as the atomic proposition that can be formulated as the environment factor to express that the occurrence of actions in the path must be true. In fact, the environment factor can be formulated in any temporal logic such as linear temporal logic (LTL) to formalize temporal properties as conditions.

We consider a simple network consisting of two parallel programs. One program sends a signal $s$ through channel $c$ while another program receives a signal from channel $c$.

The sending program specified in \eqref{f:env_sending} does not need to perceive the environment factor, meaning that it can send $s$ at any time.
\begin{align}
\label{f:env_sending}
\begin{split}
    \{ \top, \top \land s \mapsto - \} \\
    c!s \\
    \{ \top, c!s \}
\end{split}
\end{align}

The receiving program can only execute the receiving action after perceiving that the signal has been sent, which is specified in \eqref{f:env_receiving}.
\begin{align}
\label{f:env_receiving}
\begin{split}
    \{ \top, v \mapsto - \} \\
    \{ c!s, v \mapsto - \} \\
    c?v \\
    \{ c!s, c?v \land \llbracket v \rrbracket = \llbracket s \rrbracket \}
\end{split}
\end{align}

To give the proof for environment extension, we introduce \textbf{Environment Composition Rule} in \eqref{f:env_rule}.

\begin{align}
\label{f:env_rule}
\begin{split}
\infer{\{ \circledast_{i=0}^n \Upsilon_i \}~\alpha_0 \parallel ... \parallel \alpha_n~\{ \circledast_{i=0}^n \Upsilon_i' \}}
{\{ \Gamma_0, \Upsilon_0 \}~\alpha_0~\{ \Gamma_0, \Upsilon_0' \}~...~\{ \Gamma_n, \Upsilon_n \}~\alpha_n~\{ \Gamma_n, \Upsilon_n' \}} \\
\text{(Environment Composition Rule)}
\end{split}
\end{align}

For brevity, we use $\Upsilon$ to denote the conjunction of $\gamma$ and $P$. The big star notation $\circledast_i^n$ denotes consecutive separating conjunction from index $i$ to $n$.

It is notable that the inference in \textbf{Environment Composition Rule} eliminates the foreign environment naturally if we regard the proved parallel system as the highest level of specification.

\begin{figure}
\centering
\includegraphics[width=5.5cm]{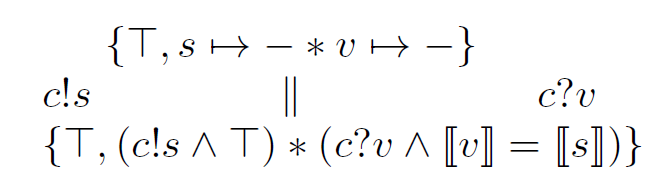}
\caption{Specification of the network in Section~\ref{sec:env}.}
\label{fig:env_spec}
\end{figure}

In the example above, we can specify two programs locally and combine the local specifications to produce a high-level specification in Fig.~\ref{fig:env_spec}. The verification can be done with \textbf{Environment Composition Rule}.

The capability of the environment perception enhances the expressiveness and extends the standard CSL into temporal formalization. The new proof rule also preserves the modularity to formalize a complex system with environment factors by permitting reasoning about programs locally.

\subsection{Node-Level Reasoning}
In decentralized systems, nodes are distributed in the network and interact with each other under specific protocols including communication protocols and execution protocols. Each node in the network perceives states of others by communication protocols regularizing the way of passing information in the network. With the information perceived from communication protocols and local states, a node enforces execution protocols with corresponding parameters. According to the scope of communication and execution protocols, it is reasonable to regard the enforcement of communication protocols as the foreign environment while the enforcement of execution protocols being the native environment.

In node-level reasoning, communications from programs on the same node and on other nodes can be distinguished with the introduction of node-level parallelism. Besides, the temporal conditions imposed by other nodes can be used as the environment factors while specifying a parallel system on a node. For instance, to do action $c!s_1$, the sending program on node $N$ needs to satisfy that another program on node $N$ has received a signal $s_0$ from another node $N'$ through channel $c$ and assigned $s_0$ to variable $v$, which can be specified as $\{ c!s_0@N' \triangleleft c?v@N, s_1 \mapsto - \}~c!s_1~\{ \top, c!s_1 \}$.

To facilitate formalizing and verifying the node and the interactions among nodes, we extend the CSL with the support of the node parallel. We introduce \textbf{Node Environment Composition Rule} in \eqref{f:node_env_rule} and \textbf{Node Composition Rule} in \eqref{f:node_com_rule}.

\begin{align}
\label{f:node_env_rule}
\begin{split}
\infer{\{ \hat{\Gamma}, \circledast_{i=0}^n \Upsilon_i \}~\alpha_0@N \parallel ... \parallel \alpha_n@N~\{ \hat{\Gamma}, \circledast_{i=0}^n \Upsilon_i' \}}
{\{ \Gamma_0, \Upsilon_0 \}~\alpha_0@N~\{ \Gamma_0, \Upsilon_0' \}~...~\{ \Gamma_n, \Upsilon_n \}~\alpha_n@N~\{ \Gamma_n, \Upsilon_n' \}} \\
\text{(Node Environment Composition Rule)}
\end{split}
\end{align}

\begin{align}
\label{f:node_com_rule}
\begin{split}
\infer{ \{ \circledast_{i=0}^n \Upsilon_i \}~\alpha_0@N_0 \parallel_N ... \parallel_N \alpha_n@N_n~\{ \circledast_{i=0}^n \Upsilon_i' \} }
{ \{ \Gamma_0, \Upsilon_0 \}~\alpha_0@N_0~\{ \Gamma_0, \Upsilon_0' \}...\{ \Gamma_n, \Upsilon_n \}~\alpha_n@N_n~\{ \Gamma_n, \Upsilon_n' \}} \\
\text{(Node Composition Rule)}
\end{split}
\end{align}

Here, $\mathfrak{N}$ denote a set of nodes. We have $N_0,.., N_n \in \mathfrak{N}$. $@$ is the ownership relation between action and node. $(\alpha, N) \in @$ denotes action $\alpha$ happens at node $N$, meaning that node $N$ has the ownership of action $\alpha$. We use the notation $\alpha @ N$ as shorthand for $(\alpha,N) \in @$. $\parallel_N$ is the notation for node-level parallel to distinguish with program-level parallel $\parallel$. For a node $N$, the foreign environment $\Gamma$ includes the temporal conditions imposed by other programs on $N$ and the temporal conditions imposed by other nodes. $\hat{\Gamma}$ denotes the foreign environment without the temporal conditions imposed by other programs on $N$.

In \textbf{Node Environment Composition Rule}, the foreign environment factors on node $N$ are composite in the inference with the persistence of the temporal conditions imposed by other nodes and the elimination of conditions imposed by other programs on $N$. Local specifications on different nodes can be combined to make a high-level specification with \textbf{Node Composition Rule}.

\section{Application}
\label{sec:app}
In this section, we present two practical applications to show how our method works with our logic including a consensus mechanism and a smart contract. These are two critical decentralized techniques widely used in decentralized systems. By formalizing and verifying them, it proves the effectiveness of our method applied in decentralized systems.

\subsection{Consensus Mechanism}
\label{sec:consensus}
We use our logic to formalize the consensus mechanism, a critical role in decentralized techniques. Hashgraph \cite{baird_swirlds_2016} is a recently developed consensus algorithm, which adopts a directed acyclic graph (DAG) structure and proves effective in permissioned blockchain.

Let us recall the data structures in Hashgraph. An event $e$ is a tuple defined in \eqref{f:def_event}.

\begin{align}
\label{f:def_event}
e \triangleq \langle \textit{TS}, \textit{TX}, \textit{SH}, \textit{OH} \, \rangle
\end{align}

Here, $\textit{TS}$ denotes a timestamp signed by the creator. $\textit{TX}$ is a set of transactions embedded into the event. $\textit{SH}$ and $\textit{OH}$ denote pointers pointing to a self-parent and an other-parent respectively. In this manner, all events associated with a set of transactions compose a DAG.

For a node $N$, we consider an immutable transaction list $\mathbf{T}$ to persist transactions accepted in the network. $\mathbf{T}$ is used to simplify the DAG structure. The acceptance is the alternative of \textit{round received}. There is a set of events $E$ associated with a set of transactions \textit{TX} on $N$. Each event is either accepted by $N$ or rejected by $N$. For brevity, we introduce two notations $\ominus$ and $\oplus$ to denote acceptance and rejection. Formally, we have definitions in \eqref{f:def_e}.

\begin{align}
\label{f:def_e}
\begin{split}
\ominus e \triangleq t \in e.\textit{TX} \implies t \notin \mathbf{T} \\
\oplus e \triangleq t \in e.\textit{TX} \implies t \in \mathbf{T} \\
\ominus E \triangleq \forall e \in E: t \in e.\textit{TX} \implies t \notin \mathbf{T} \\
\oplus E \triangleq \exists e \in E: t \in e.\textit{TX} \implies t \in \mathbf{T}
\end{split}
\end{align}

In this paper, we consider a small network to illustrate our methodology of reasoning about the Hashgraph mechanism in an outline. With the support of modular reasoning, our logic can reason about nodes and programs on nodes separately.

There are four nodes $N_0$, $N_1$, $N_2$, $N_3$ in the network. Each node is deployed with programs enforcing the Hashgraph consensus mechanism. Generally, there must be at least one program for broadcasting events to other nodes and one program for receiving events from other nodes. All programs run in a concurrent manner.

\begin{figure}
\centering
\includegraphics[width=\textwidth]{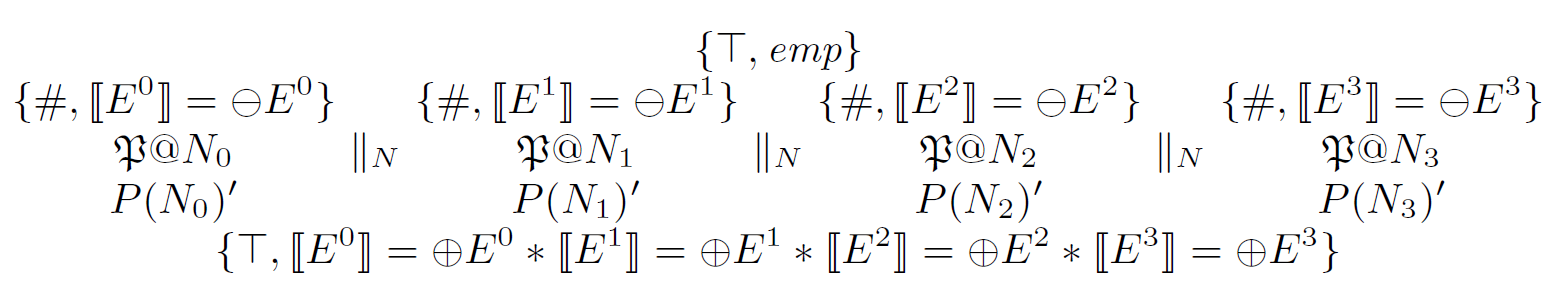}
\caption{Specification of the event consensus in Hashgraph network.}
\label{fig:hashgraph_spec}
\end{figure}

We specify the outline of the consensus mechanism for a set of transactions $\textit{TX}$ in Fig.~\ref{fig:hashgraph_spec}, where $\#$ denotes the occurred action that receives transactions from a client. $E^i$ is a set of events created on node $N_i$. Let $e^i_j$ denotes the $j$th event on node $N_i$ where $j \in \mathbb{N}$, we have $\forall e^i_j \in E^i \setminus \{ e^i_0 \}: e^i_j.\textit{SH} = \mathcal{H}(e^i_{j-1})$ where $\mathcal{H}$ is the hash function. $P(N_i)'$ is the post-condition of node $N_i$, which specifies the termination condition of the consensus on $\textit{TX}$. For instance, we have $P(N_0)'$ as follows:

$P(N_0)' = \{ \Diamond \textit{RR}(E^0)@N_1 \land \Diamond \textit{RR}(E^0)@N_2 \land \Diamond \textit{RR}(E^0)@N_3), \Diamond \textit{RR}(E^0)@N_0 \land \llbracket E^0 \rrbracket = \oplus E^0 \}$,

where $\textit{RR}(E)$ denotes the action \textit{round received}, meaning that an event $e \in E$ is accepted in some round. In this application, we use LTL to formulate the environment factor. Here, $\Diamond$ is a temporal modality denoting \textit{eventually}.

This is a high-level specification of the whole network to verify that all nodes will eventually get the consistent $\mathbf{T}$. With the support of \textbf{Node Composition Rule}, we can break down the specification and focus on specifying and verifying programs on nodes separately. On each node, there is a parallel system containing a set of programs, which can also be formalized separately.

Besides, We can combine several highly coupled programs as a module. For instance, \textit{famous witness election} is a critical module to make the Hashgraph algorithm converge. Recall the concept of \textit{witness} in Hashgraph that is the first event a node creates in each round. Each node will decide its witness events after executing action \textit{round created}. The election is for witnesses to vote on a witness before the current round to determine whether it is famous. In fact, the vote is virtual, which means that there is not an actual vote based on communications formed in the network. The voting process is enforced on the local environment. Here, for a vote function, we can specify it as \eqref{f:witness}.

\begin{align}
\label{f:witness}
\{ \top, \Diamond \textit{See}(e,e') \land \textit{IsWitness}(e) \land \textit{IsWitness}(e') \}~\textit{Vote}(e,e')~\{ \top, \textit{IsFamous}(e,e') \}
\end{align}

Here, the specification has the meaning that if a witness $e$ can \textit{see} another witness $e'$, then $e$ votes $e'$ as a famous witness and the ballot is signed with $e$.

Besides, vote collection is also enforced on the local environment in the round after voting. Each node can make the election by itself without synchronous interactions with others based on a directed acyclic graph that is a data structure that can be trivially formalized in the standard CSL.

In this manner, our logic can effectively formalize such a module with modular reasoning and other critical modules such as \textit{see} and \textit{strongly see} with the support of environment perception.

\subsection{Smart Contract}
\label{sec:smart_contract}
We follow the unitarity principle to unify the specification and verification of decentralized systems at different abstraction levels with our logic. Besides complex protocols such as consensus mechanisms, our logic can also specify and verify the decentralized applications developed by smart contracts.

We take a simple example of the smart contract slightly modified from \cite{bhargavan_formal_2016}, which is similar to the behavior of TheDAO attack. It defines a contract named \textit{MyBank} with three functions \textit{Deposit}() for depositing money, \textit{Withdraw}(amount::Integer) for withdrawing money, and \textit{Balance}() for looking up the balance. Another contract named \textit{Malicious} behaves maliciously with two evil functions \textit{Drain}() to trigger the malicious behavior and a fallback method, which is shown in Listing~\ref{lst:malicious}.

\begin{lstlisting}[caption={Pseudocode of Malicious contract}, language=Java,mathescape=true, numbers=left, label={lst:malicious},float=*]
contract Malicious {
    ...
    function Drain() {
        myBank.Deposit.value(amount);
        myBank.Withdraw.value(amount);
    }

    function () {
        myBank.Withdraw.value(amount);
    }
    ...
}
\end{lstlisting}

With our logic, we can specify the contract \textit{MyBank} in Listing~\ref{lst:mybank}.

\begin{lstlisting}[caption={Pseudocode of MyBank contract}, language=Java,mathescape=true, numbers=left, label={lst:mybank},float=*]
contract MyBank {
    ...
    $\{ c?\textit{msg}, \textit{balances}[msg.sender] \mapsto - \}$
    function Deposit() {
        $\{ \top, \llbracket \textit{balances}[msg.sender] \rrbracket = n \}$
        balances[msg.sender] += msg.value;
        $\{ \top,  \llbracket \textit{balances}[msg.sender] \rrbracket = n +\textit{msg.value} \}$
    }
    $\{ \top,  \llbracket \textit{balances}[msg.sender] \rrbracket = n +\textit{msg.value} \}$
    
    $\{ c?\textit{msg} \land c?\textit{amount}, \textit{balances}[msg.sender] \mapsto - \}$
    function Withdraw(amount::Integer) {
        $\{ \top, \llbracket \textit{balances}[msg.sender] \rrbracket = n \land n \geq \textit{amount} \}$
        if (balances[msg.sender] $\geq$ amount) {
            msg.sender.call.value(amount);
            balances[msg.sender] -= amount;
        }
        $\{ \top, \llbracket \textit{balances}[msg.sender] \rrbracket = n-amount] \}$
    }
    $\{ \top, \llbracket \textit{balances}[msg.sender] \rrbracket = n-amount] \}$
    ...
}
\end{lstlisting}

Then, we consider a concurrent specification and introduce a resource invariant $\textit{RI}_{b}$ where resource $b$ is the balance. We have $\textit{RI}_{b} \triangleq b \geq 0$. The function \textit{Withdraw} is specified in \eqref{f:withdraw} from a high-level while \textit{Deposit} is specified in the same manner.

\begin{align}
\label{f:withdraw}
\begin{split}
\{ c?\textit{msg} \land c?a, \textit{RI}_b \land \llbracket b \rrbracket = n \} \\
\textit{Withdraw}(a)' \\
\{ \top, \textit{RI}_b \land \llbracket b \rrbracket = n - a \}
\end{split}
\end{align}

\begin{figure}
\centering
\includegraphics[width=\textwidth]{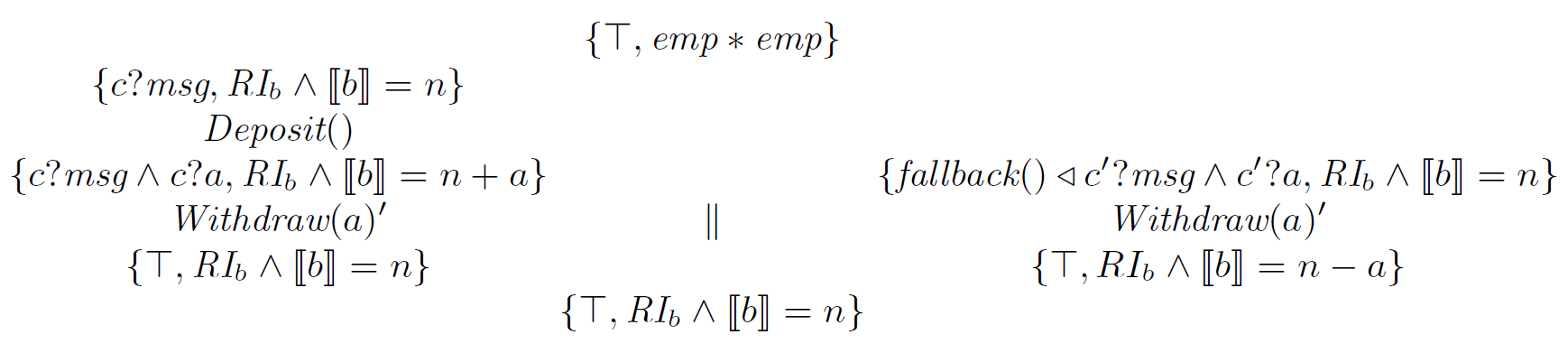}
\caption{Specification of the parallel smart contract system.}
\label{fig:sc_spec}
\end{figure}

The function \textit{msg.sender.call.value} in List~\ref{lst:mybank} transfers control back to contract \textit{Malicious} by invoking the fallback method. In the fallback method, a reentrant call \textit{Withdraw}() is triggered in another thread or process of $\textit{MyBank}'$ as the shadow of $\textit{MyBank}$, which can be specified in Fig.~\ref{fig:sc_spec}.

We can obtain that the specification of this parallel smart contract system cannot be proved in any inference rules while preserving the resource invariant and satisfying the post-condition if both $\textit{MyBank}$ contract program and its shadow program $\textit{MyBank}'$ are verified separately. Therefore, the design flaw is located. In this manner, our logic can prove the correctness of smart contracts and assist in finding design flaws in decentralized applications.

\section{Discussion}
\label{sec:discuss}
The core of our method of formalizing and verifying decentralized systems is our extended concurrent separation logic. Our logic presents rich expressiveness with the encapsulation of communication formalization and the capability of environment perception. It also has great modularity with the node-level reasoning feature to enable formalizing decentralized systems from memory level to node level. Furthermore, the formalization and application at different abstraction levels present unitarity with unified forms.

In the meanwhile, our logic has compatibility, which allows the interpretation from our logic to the standard CSL. We have proved the soundness of our logic by the general structure of \cite{vafeiadis_concurrent_2011}. We have also mechanized our logic by an interpreter together with a mechanized CSL prover implemented in the Coq proof assistant to implement the applications in Section~\ref{sec:app}. However, it is not dependable to reduce our logic to the CSL for automated reasoning by the interpreter. We plan to mechanize our logic directly with the proof assistant.

We do not address formulating temporal properties as temporal conditions in the environment factors in this paper. However, it is straightforward to formulate them in temporal logic. For instance, we can express temporal properties with LTL formulae such as $\{ \phi, \top \}~\alpha~\{ \top, \top \}$ to add the constraint for the occurrence of $\alpha$ with the LTL formula $\phi$.

\section{Conclusion}
\label{sec:conclusion}
Reasoning about decentralized systems during the development is indispensable to ensure the correctness and system properties since it is a significant challenge to establish trust without central entities in untrustworthy distributed environments. In this paper, we have proposed a novel method of formalizing and verifying decentralized systems with an extended concurrent separation logic enhanced by three novel features: communication encapsulation, environment perception, and node-level reasoning. Besides, we have applied our method to reason about the consensus mechanism to prove the effectiveness of reasoning about complex protocols with the support of new features. Furthermore, we have demonstrated the capability of formalizing and verifying smart contracts with an instance to locate the design flaw. Particularly, we follow the unitarity principle and compatibility principle to facilitate the implementation. We plan to optimize the mechanization of our logic directly with the proof assistant. Formalizing and verifying a more complicated decentralized system such as a blockchain platform is also on our schedule.

%
%
%
\bibliographystyle{splncs04}
\bibliography{formalizeds}

\end{document}